%% file: HIN-16-011_temp.tex
\pdfoutput=1

\documentclass[11pt,twoside,a4paper,cmspaper,final,collab]{cms-tdr}
\def\svnVersion{429841P}\def\cmsCernNoTag{CERN-EP-2017-076}\def\cmsCernDate{\today}\def\cmsMessage{Published in Physical Review Letters as \href{http://dx.doi.org/10.1103/PhysRevLett.119.152301}{\doi{10.1103/PhysRevLett.119.152301.}}}

\begin{document}\cmsNoteHeader{HIN-16-011}

\hyphenation{had-ron-i-za-tion}
\hyphenation{cal-or-i-me-ter}
\hyphenation{de-vices}
\RCS$HeadURL: svn+ssh://svn.cern.ch/reps/tdr2/papers/HIN-16-011/trunk/HIN-16-011.tex $
\RCS$Id: HIN-16-011.tex 423381 2017-09-01 14:20:05Z alverson $
\newlength\cmsFigWidth
\ifthenelse{\boolean{cms@external}}{\setlength\cmsFigWidth{0.98\columnwidth}}{\setlength\cmsFigWidth{0.69\textwidth}}
\ifthenelse{\boolean{cms@external}}{\providecommand{\cmsLeft}{top\xspace}}{\providecommand{\cmsLeft}{left\xspace}}
\ifthenelse{\boolean{cms@external}}{\providecommand{\cmsRight}{bottom\xspace}}{\providecommand{\cmsRight}{right\xspace}}
\newcommand{\sqrts}{ \ensuremath{\sqrt{s}}\xspace}
\newcommand{\sqrtsNN}{ \ensuremath{\sqrt{\smash[b]{s_{_{\mathrm{NN}}}}}}\xspace}
\newcommand{\Bplusminusdecay}{\ensuremath{\PBpm\to\PJGy~\PKpm\to\Pgmp\Pgmm\PKpm}\xspace}
\newcommand{\Bjpsixdecay}{\ensuremath{\PB\to\PJGy~\cmsSymbolFace{X}}\xspace}
\newcommand{\RAA}{\ensuremath{R_{\mathrm{AA}}}\xspace}
\newcommand{\TAA}{\ensuremath{T_{\mathrm{AA}}}\xspace}
\newcommand{\pPb}{\ensuremath{\Pp\mathrm{Pb}}\xspace}
\newcommand{\PbPb}{\ensuremath{\mathrm{PbPb}}\xspace}
\newcommand{\Pb}{\ensuremath{\mathrm{Pb}}\xspace}
\newcommand{\pp}{\Pp\Pp\xspace}

\providecommand{\mbinv} {\mbox{\ensuremath{\,\text{mb}^\text{$-$1}}}\xspace}
\providecommand{\HYDJET}{\textsc{hydjet}\xspace}

\cmsNoteHeader{HIN-16-011}

\title{\texorpdfstring{Measurement of the $\PBpm$ meson nuclear modification factor in \PbPb collisions at $\sqrtsNN=5.02\TeV$}{Measurement of B+/- mesons nuclear modification factor in PbPb collisions at sqrt(s[NN]) = 5.02 TeV}}

\date{\today}

\abstract{
The differential production cross sections of $\PBpm$ mesons are measured via the exclusive decay channels \Bplusminusdecay as a function of transverse momentum in \pp and \PbPb collisions at a center-of-mass energy $\sqrtsNN=5.02\TeV$ per nucleon pair with the CMS detector at the LHC.
The \pp (\PbPb) dataset used for this analysis corresponds to an integrated luminosity of 28.0\pbinv (351\mubinv).
The measurement is performed in the $\PBpm$ meson transverse momentum range of 7 to 50\GeVc, in the rapidity interval $\abs{y}<2.4$. In this kinematic range, a strong suppression of the production cross section by about a factor of two is observed in the \PbPb system in comparison to the expectation from \pp reference data. These results are found to be roughly compatible with theoretical calculations incorporating beauty quark diffusion and energy loss in a quark-gluon plasma.
}

\hypersetup{%
pdfauthor={CMS Collaboration},%
pdftitle={Measurement of B+/- meson differential production cross sections in pp and PbPb collisions at sqrt(s[NN]) = 5.02 TeV},%
pdfsubject={CMS},%
pdfkeywords={physics, dimuons, proton Lead, charmonia, suppression, quark gluon plasma, shadowing, B meson, open heavy-flavor}}

\maketitle
Relativistic heavy ion collisions allow the study of quantum chromodynamics (QCD) at high energy density.
Under such extreme conditions, a state consisting of deconfined quarks and gluons, the quark-gluon plasma (QGP)~\cite{QGP1,QGP2}, is predicted by lattice QCD calculations~\cite{Karsch:2003jg}.
Hard-scattered partons are expected to lose energy via elastic collisions and medium-induced gluon radiation as they traverse the QGP. This phenomenon, known as jet quenching~\cite{Eloss1,Baier:2000mf,Chatrchyan:2011sx,Aad:2010bu}, results in the suppression of the yield of high transverse momentum (\pt) hadrons, compared to the  expectation based on proton-proton (pp) data, in which the outgoing partons traverse the QCD vacuum. Measurements of the jet quenching dependence on the type of initiating parton (both quark \vs gluon and light \vs heavy quarks) are key to constrain the QGP properties~\cite{Dokshitzer:2001zm,Armesto:2003jh,Wicks:2007am,Zhang:2003wk,Adil:2006ra}.

The production of $\PB$~mesons was studied at the Large Hadron Collider (LHC) in \pp collisions at center-of-mass energies of $\sqrts = 7$\TeV~\cite{CMSBmesonpp,Chatrchyan:2011pw,Chatrchyan:2011vh,ATLAS:2013cia,LHCb:2013JHEP,Aaij:2014hla,Aaij:2012dd}, 8\TeV~\cite{Aaij:2015fea,Aaij:2014ija} and 13\TeV~\cite{Khachatryan:2016csy} over wide \pt and rapidity ($y$) intervals, and in proton-lead ($\pPb$) collisions at a center-of-mass energy per nucleon pair $\sqrtsNN = 5.02$\TeV~\cite{Khachatryan:2015uja}.
The CMS Collaboration also measured the nonprompt (\ie from decays of $\PQb$ hadrons) $\PJGy$ meson production in lead-lead ($\PbPb$) and \pp collisions at $\sqrtsNN = 2.76$\TeV~\cite{CMSNonPromptJpsi}. For nonprompt $\PJGy$, a strong suppression was observed in the nuclear modification factor \RAA, the ratio of the nonprompt \PJGy cross section in \PbPb collisions with respect to that in \pp collisions scaled by the number of binary nucleon-nucleon (NN) collisions. In this Letter, we extend the study of heavy-quark production by performing the first measurement of exclusive $\PBpm$ mesons decays in $\PbPb$ collisions. This provides direct information about the b hadron kinematics and flavor content, compared to the measurements of nonprompt $\PJGy$, which are decay products of various beauty mesons and baryons.

The $\PBpm$ mesons are measured in the interval $\abs{y}<2.4$ and in five \pt bins ($[7,10]$, $[10,15]$, $[15,20]$, $[20,30]$, $[30,50]$\GeVc), via the reconstruction of the decay channels \Bplusminusdecay, which have the branching fraction $\mathcal{B} = (6.12 \pm 0.19) \times 10^{-5}$~\cite{pdg:2016}. Throughout the paper, unless otherwise specified, the $y$ and \pt variables given are those of the $\PBpm$ mesons. This analysis does not distinguish between the charge conjugates.

The central feature of the CMS detector is a superconducting solenoid which provides a magnetic field of 3.8\unit{T}. Within the solenoid volume are a silicon tracker which measures charged particles within the pseudorapidity range $\abs{\eta} < 2.5$, a lead tungstate crystal electromagnetic calorimeter, and a brass and scintillator hadron calorimeter. For typical particles of $1 < \pt < 10$\GeVc and $\abs{\eta} < 1.4$, the track resolutions are typically 1.5\% in \pt and 25--90 (45--150)\mum in the transverse (longitudinal) impact parameter \cite{TRK-11-001}. Muons are measured in the range $\abs{\eta} < 2.4$, with detection planes made using three technologies: drift tubes, cathode strip chambers, and resistive-plate chambers. The muon reconstruction algorithm starts by finding tracks in the muon detectors, which are then fitted together with tracks reconstructed in the silicon tracker to form "global muons". Matching muons to tracks measured in the silicon tracker results in a relative \pt resolution for muons with $20 < \pt < 100$\GeVc of 1.3--2.0\% in the barrel ($\abs{\eta} < 1.2$) and better than 6\% in the endcaps ($1.6 < \abs{\eta} < 2.4$).
For muons with higher \pt up to 1\TeVc, the \pt resolution in the barrel is better than 10\%~\cite{Chatrchyan:2012xi}.  The hadron forward (HF) calorimeter uses steel as an absorber and quartz fibers as the sensitive material. The two halves of the HF are located 11.2\unit{m} away from the interaction point, one on each end, providing together coverage in the range $3.0 < \abs{\eta} < 5.2$. In this analysis, the HF information is used for performing an offline event selection. A detailed description of the CMS experiment and coordinate system can be found in Ref.~\cite{bib_CMS}.

For the decay channel measured in this analysis, the background consists primarily of two sources.
A combinatorial background originates from randomly pairing a $\PJGy$ with an unrelated charged particle. This gives rise to a falling contribution in the invariant mass spectrum.
A heightened background in the invariant mass region below 5.4\GeVcc is also present, which corresponds to partially reconstructed b hadron decays from processes other than the one of interest.
As an example, a heightened structure can be created by ${\text{B}}^{0} \to \PJGy~\PKst^0 \to \mu^{+}\mu^{-}\PKp\Pgpm$ ($\PaBz \to \PJGy~\overline{\PKst^0} \to \mu^{+}~\mu^{-}\PKm\Pgpp$) decays in which one decay product is lost, resulting in a $\PBp$ (\PBm) candidate.
Several Monte Carlo (MC) simulated event samples are used to evaluate background components, signal efficiencies and detector acceptance corrections. This includes samples containing only the $\PBpm$ mesons decays channels being measured, and samples with inclusive (prompt and nonprompt) $\PJGy$ mesons.
Proton-proton collisions are generated with \PYTHIA 8~\cite{Sjostrand:2014zea} tune CUETP8M1~\cite{Khachatryan:2015pea} and propagated through the CMS detector using the \GEANTfour package~\cite{geant4}.
The decay of the $\PB$~mesons is modeled with the \EVTGEN 1.3.0~\cite{evtgen}, and final-state photon radiation in the $\PB$ decays is simulated with \PHOTOS 2.0~\cite{Barberio:1990ms}. For the $\PbPb$ MC samples, each \PYTHIA 8 event is embedded into a $\PbPb$ collision event generated with {\HYDJET}~1.8~\cite{Lokhtin:2005px}, which is tuned to reproduce global event properties, such as the charged-hadron \pt spectrum and particle multiplicity.

Events were collected with the same trigger during the \pp and \PbPb data taking, requiring the presence of two muon candidates, with no explicit momentum threshold. For the offline analysis, events have to pass a set of selection criteria designed to reject events from background processes
(beam-gas collisions and beam scraping events) as described in Ref.~\cite{Khachatryan:2016odn}. Events are required to have
at least one reconstructed primary interaction vertex with a distance from the center of the nominal interaction region of less than 15\cm along the beam axis. In \PbPb collisions, the
shapes of the clusters in the pixel detector have to be
compatible with those expected from particles produced by a \PbPb
collision~\cite{Khachatryan:2010xs}. The \PbPb collision event is also required to have at least
three towers in each of the HF detectors with energy deposits of more
than 3\GeV per tower. These criteria select $(99\pm2)$\% of inelastic hadronic \PbPb collisions. Selection efficiencies higher
than 100\% are possible, reflecting the possible presence of ultra-peripheral (i.e. nonhadronic) collisions in the selected
event sample.
The \PbPb sample corresponds to an integrated luminosity of approximately 351\mubinv. This value is indicative only, as the \PbPb yield is normalized by the total number of minimum-bias events sampled, $N_{\text{MB}}$. The \pp data set corresponds to an integrated luminosity of 28.0\pbinv, which is known to an accuracy of 2.3\% from the uncertainty in the calibration based on a van der Meer scan~\cite{CMS-PAS-LUM-16-001}.

Kinematic limits are imposed on the single muons so that their reconstruction efficiency stays above 10\%. These limits are $\pt^{\mu}>3.5\GeVc\enspace \text{ for } \abs{\eta^{\mu}}<1.2$, $\pt^{\mu}>1.8\GeVc\text{ for }2.1\le\abs{\eta^{\mu}}<2.4$, and linearly interpolated in the intermediate $\abs{\eta^{\mu}}$ region. The muons are also required to match the muons that triggered the event online, and pass selection criteria optimized for low \pt (the so-called \textit{soft selection}~\cite{Chatrchyan:2012xi}). Two muons of opposite charge with an invariant mass within $150\MeVcc$ of the world-average $\PJGy$ meson mass~\cite{pdg:2016} are selected to reconstruct a $\PJGy$ candidate, with a mass resolution of typically 18--55\MeVcc, degrading as a function of the dimuon rapidity and \pt. Opposite-sign muon pairs are fitted with a common vertex constraint and are kept if the $\chi^2$ probability of the fit is greater than 1\%, lowering the background from charm- and beauty-hadron semileptonic decays. Each $\PB$~meson candidate is formed from the combination of a $\PJGy$ candidate with a charged-particle track, which are required to pass standard selections described in Ref.~\cite{Khachatryan:2016odn}. Without using particle identification, assumptions need to be made about the masses of the charged particles. In calculating the mass of the $\PBpm$ candidates, the single charged particle is always assumed to have the mass of a charged kaon, and the muon pair is assumed to have the mass of a $\PJGy$ meson.
A single-track low-$\pt$ threshold of 0.5\GeVc for \pp collisions and 0.8\GeVc for \PbPb collisions is applied to reduce the combinatorial background, which is further minimized by additional selection criteria.
In particular, $\PBpm$ candidates are selected according to the $\chi^{2}$ probability of their decay vertex (the probability for the muon tracks from the $\PJGy$ meson decay and the other charged track to point to a common vertex), the three-dimensional (3D) flight distance (normalized by its uncertainty) between the primary and decay vertices, and the pointing angle, which is defined as the angle between the line segment connecting the primary and decay vertices and the momentum vector of the $\PBpm$ meson in the plane transverse to the beam direction. The selection is optimized in each \pt bin, separately for \pp and \PbPb results, using a multivariate technique that utilizes the genetics algorithm~\cite{Hocker:2007ht}, in order to maximize the statistical significance of the $\PBpm$ mesons signal.

\begin{figure*}[tb]
\centering
\includegraphics[width=.49\textwidth]{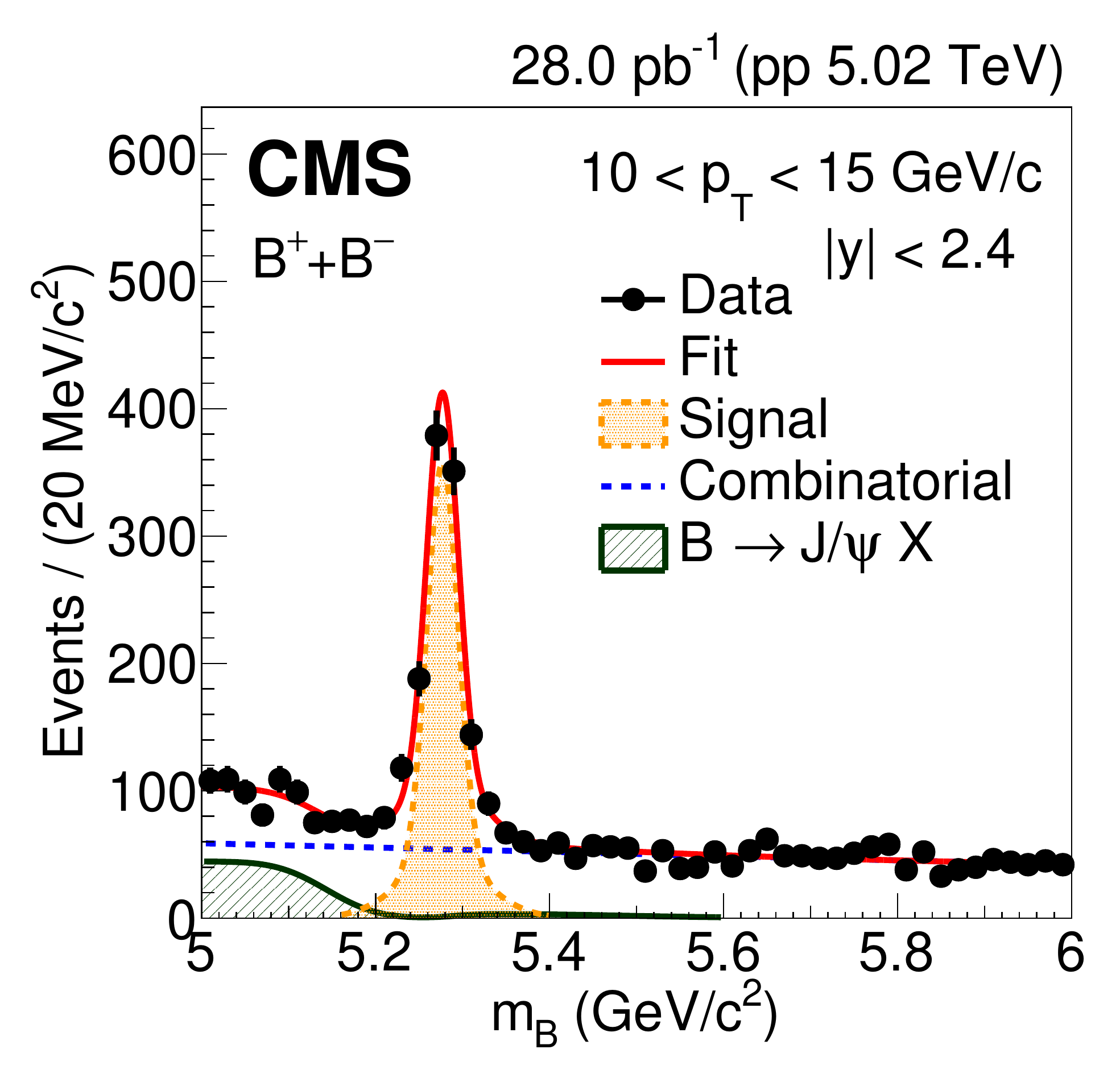}
\includegraphics[width=.49\textwidth]{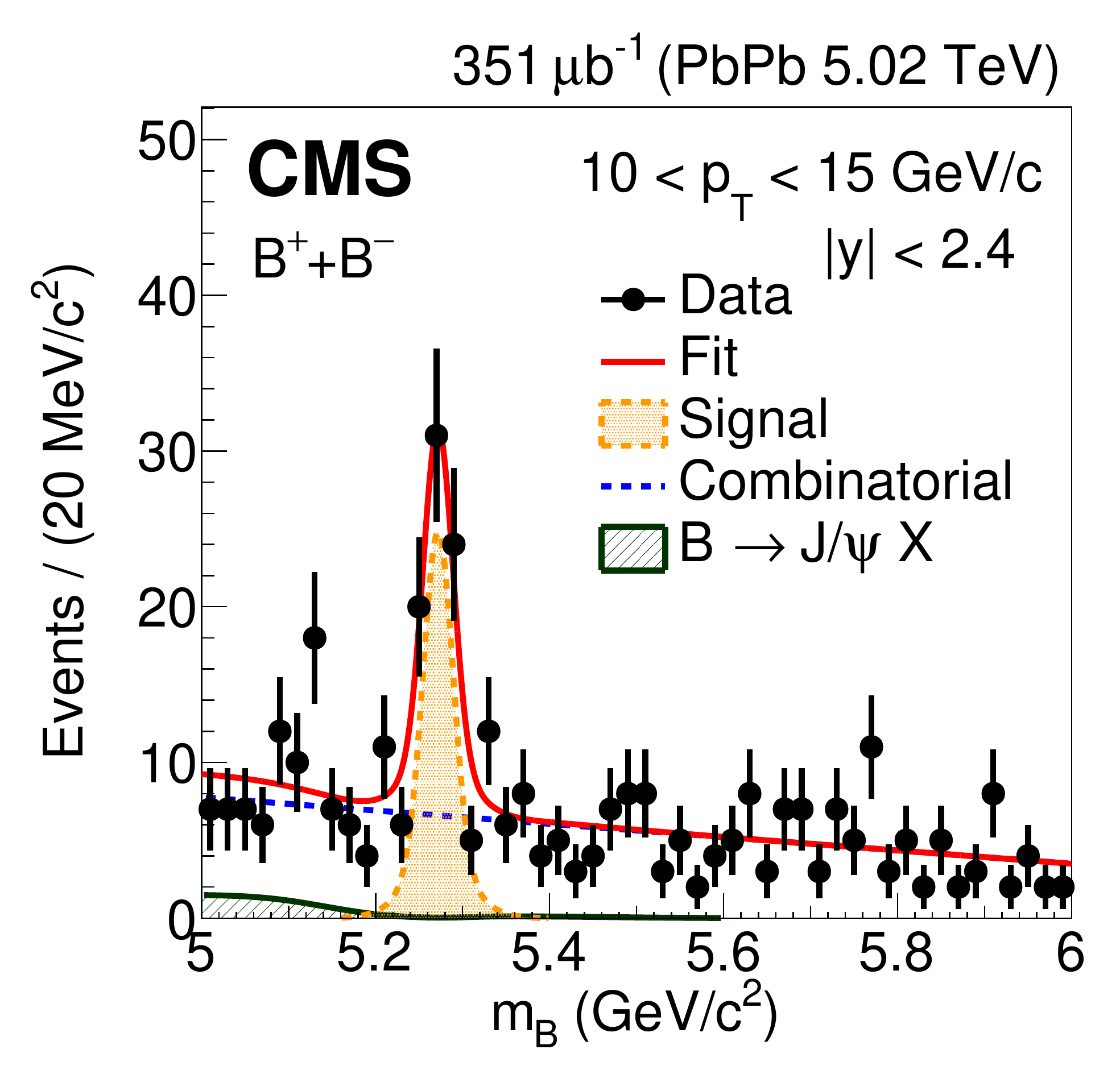}
\caption{Invariant mass distributions of $\PBpm$ candidates in \pp (left) and \PbPb (right) collisions measured in $\abs{y}<2.4$ and in the \pt region 10--15\GeVc.}
\label{fig:rawYieldsBmeson}

\end{figure*}

The raw yields of $\PBpm$ mesons in \pp and \PbPb collisions are extracted using a binned maximum likelihood fit to the $\PBpm$ mesons invariant mass distributions in the mass range 5--6$\GeVcc$. The estimation of the statistical uncertainties of the fitted
raw yields is based on the second derivatives of the
negative log-likelihood function. Examples of fits to the invariant mass distributions in \pp and \PbPb collisions are shown in Fig.~\ref{fig:rawYieldsBmeson}, for the $\pt$ region 10--15\GeVc.
The signal shape is modeled by two Gaussian functions with a common mean, a free parameter of the fit, and different widths determined from MC simulation for each \pt bin, individually for the \pp and \PbPb results. The relative contribution of the two Gaussian functions to the signal yield is also fixed at the value given by the MC sample.
The combinatorial background is modeled by a first-order polynomial as determined by studies of the inclusive $\PJGy$ MC sample. The peaking background, labeled $\Bjpsixdecay$ in Fig.~\ref{fig:rawYieldsBmeson}, is studied with the embedded MC sample including all $\PB$~meson decays into final states with a $\PJGy$ meson and found to be well described by the superposition of a double-sided Gaussian function and an error function. The shape is determined from a fit of the MC sample with all parameters free. The resulting functional form, with the overall normalization left floating, is included in the global fit function.

The differential cross section for $\PBpm$ production in $\abs{y}<2.4$ is computed in each \pt interval according to
\begin{linenomath}
\begin{equation}
 \left.\frac{{\rd}\sigma^{\PBpm}}{{\rd}\pt}\right|_{\abs{y}<2.4} =
 \frac{1}{2} \frac{1}{\mathcal{B} \, \mathcal{L}} \frac{1}{\Delta\pt} \left.\frac{N^{(\PBp+\PBm)}_{\rm pp}(\pt)}{\alpha_{\rm pp}(\pt) \, \epsilon_{\rm pp}(\pt)}\right|_{\abs{y}<2.4}
 \label{eq:crosssectionDpPb}
\end{equation}
\end{linenomath}
for \pp data, and for \PbPb data according to
\ifthenelse{\boolean{cms@external}}{
\begin{multline}
 \left. \frac{1}{\TAA}\frac{{\rd}N^{\PBpm}_\PbPb}{{\rd}\pt}\right|_{\abs{y}<2.4} = \frac{1}{2} \frac{1}{\mathcal{B} \, N_{\text{MB}} \, \TAA} \frac{1}{\Delta\pt} \\
 \times\left.\frac{N^{(\PBp+\PBm)}_\PbPb(\pt)}{\alpha_\PbPb(\pt) \, \epsilon_\PbPb(\pt)}\right|_{\abs{y}<2.4} .
 \label{eq:PbPbYield}
\end{multline}
}{
\begin{equation}
 \left. \frac{1}{\TAA}\frac{{\rd}N^{\PBpm}_\PbPb}{{\rd}\pt}\right|_{\abs{y}<2.4} = \frac{1}{2} \frac{1}{\mathcal{B} \, N_{\text{MB}} \, \TAA} \frac{1}{\Delta\pt} \left.\frac{N^{(\PBp+\PBm)}_\PbPb(\pt)}{\alpha_\PbPb(\pt) \, \epsilon_\PbPb(\pt)}\right|_{\abs{y}<2.4} .
 \label{eq:PbPbYield}
\end{equation}
}
The $N^{(\PBp+\PBm)}_{\pp, \PbPb}$ is the raw signal yield extracted in each \pt interval of width $\Delta \pt$, $(\alpha \, \epsilon)_{\pp, \PbPb}$ represents the corresponding acceptance times efficiency, and $\mathcal{B}$ is the branching fraction of the decay chain. For the \pp cross section, $\mathcal{L}$ represents the integrated luminosity. For the \PbPb cross section, the \TAA is the nuclear overlap function~\cite{Miller:2007ri}, equal to the number of NN binary collisions divided by the NN total inelastic cross section, and which can be interpreted as the NN-equivalent integrated luminosity per heavy ion collision.

The \TAA value for inclusive \PbPb collisions at $\sqrtsNN=5.02$\TeV is 5.61\mbinv, as estimated from a Monte Carlo Glauber model~\cite{Miller:2007ri, Khachatryan:2016odn}.

Assuming that in the kinematic region accessible by the present measurement $\PBp$and $\PBm$ production cross sections are equal, the factor 1/2 accounts for the fact that the yields are measured for particles and antiparticles added together, but the cross section is given for one species only.

\begin{figure*}[tb]
\centering
\includegraphics[width=.45\textwidth]{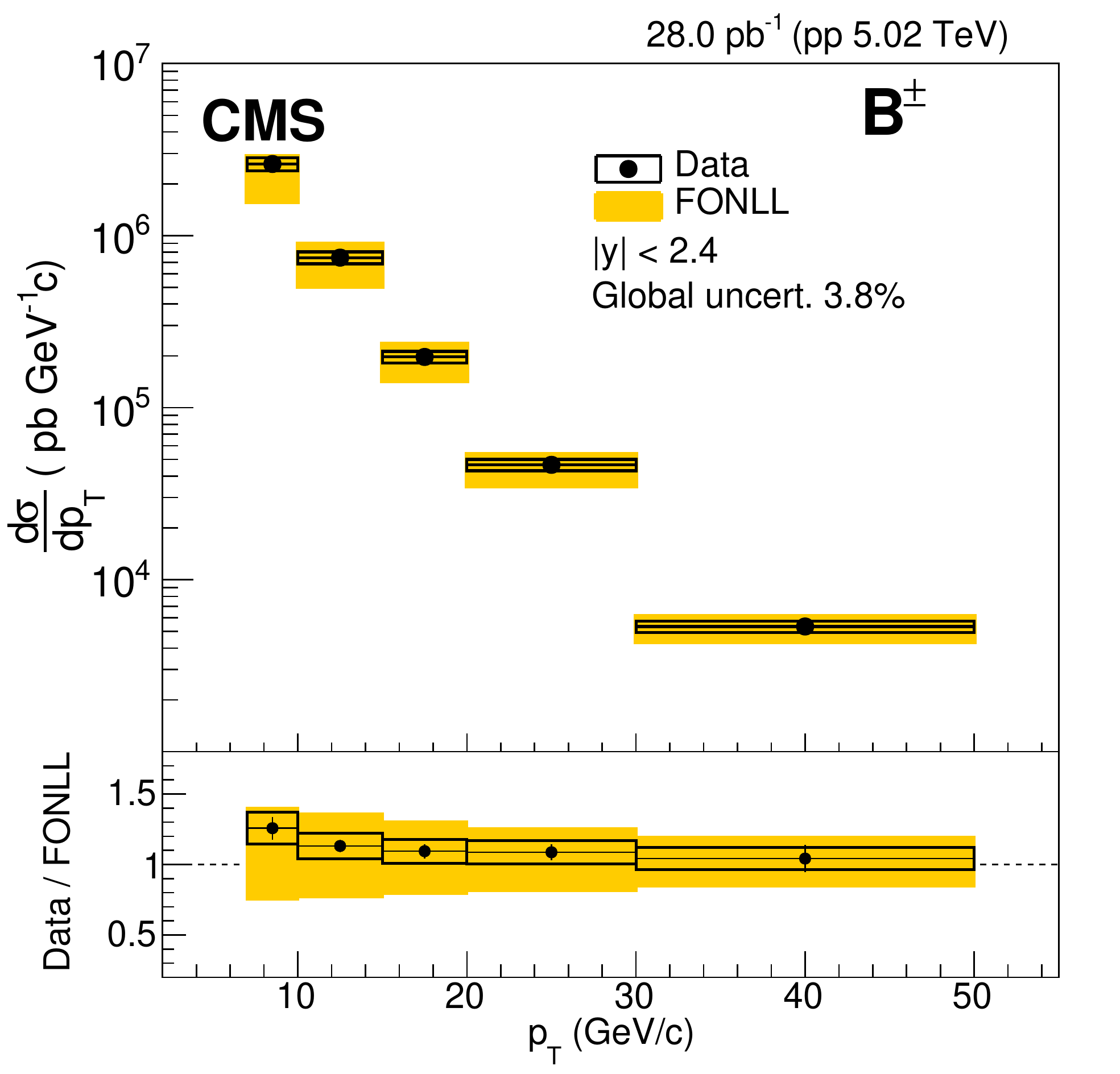}
\includegraphics[width=.45\textwidth]{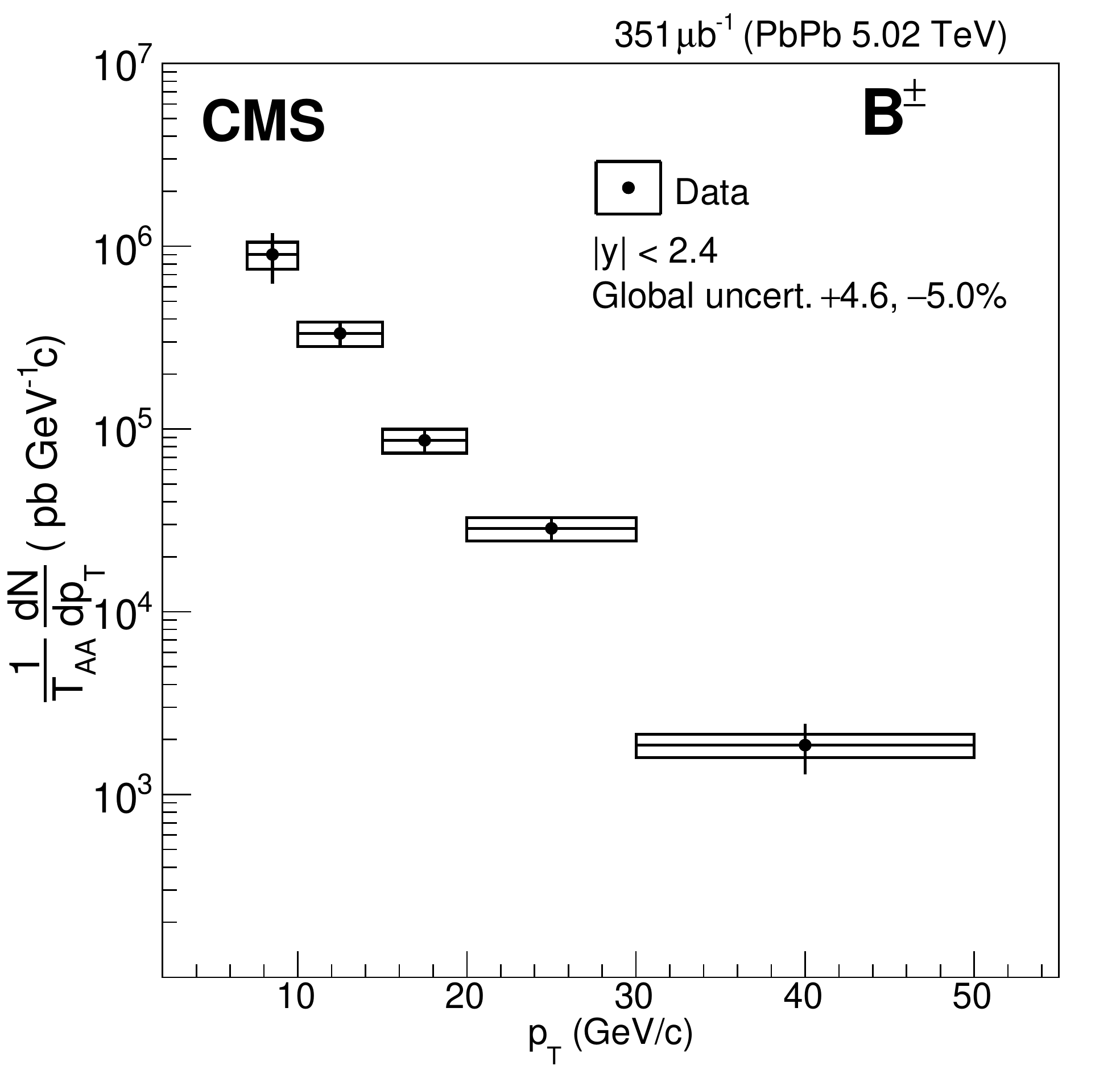}
\caption{
The \pt-differential production cross section of \PBpm in \pp (left) and \PbPb (right) collisions at $\sqrts=5.02\TeV$. The vertical bars (boxes) correspond to statistical (systematic) uncertainties.
The systematic uncertainty boxes here include both the correlated and uncorrelated contributions added in quadrature. The global systematic uncertainty, listed in the legend and not included in the point-to-point uncertainties. For the \pp cross section, they comprise the uncertainties in the integrated luminosity measurement and in the branching fraction $\mathcal{B}$. For the \PbPb cross section, they comprise the uncertainties in \TAA, $N_{\text{MB}}$, and $\mathcal{B}$.
The \pp cross section is compared to FONLL calculations~\cite{FONLLcharmbottomPP1,FONLLcharmbottomPP2,FONLLcharmbottomPP3} represented by the colored boxes with the heights indicating the theoretical uncertainty.}
\label{fig:crosssections}
\end{figure*}

The cross sections are affected by several sources of systematic uncertainties arising from the signal extraction, corrections, $\mathcal{B}$, $\mathcal{L}$ or \TAA determination.
The uncertainty of the modeling of the signal and background shapes (2.9\% and 2.6\% for \pp and \PbPb cases, respectively) is evaluated on the \pt integrated bin, by varying the probability distribution functions used to describe the signal and background distributions. As an alternative combinatorial background shape, an exponential function, and also second- and a third-order polynomials are used. The uncertainty of the signal modeling is evaluated by considering two fit variations: (i) leaving the width parameters free and (ii) using a sum of three Gaussian functions with common mean. The maximum of the signal variations is added in quadrature to the maximum of all the background variations, and  propagated as the systematic uncertainty.

The systematic uncertainty due to the selection of the $\PB$~meson candidates (3.8\% for \pp and 12.0\% for \PbPb collisions) is estimated, in the \pt integrated bin, from several variations of the selection value for each of the following: $\chi^{2}$ probability of the decay vertex, the 3D flight distance, the pointing angle, the track \pt, the track $\eta$, and the choice of the algorithm in the multivariate analysis. In each case, a systematic uncertainty is estimated from all variations, as the maximum of 1 minus the ratio of the selection efficiencies (the ratio of the nominal yield and the yield after applying the modified selection) estimated in data and simulation. The total uncertainty for the selection of the $\PB$~meson candidates is the quadratic sum of the individual contributions from the six settings.

The bin-by-bin systematic uncertainties associated with the acceptance correction (0.1\% to 0.4\%) are estimated by varying the shape of the generated $\PBpm$~meson \pt and $y$ spectra within limits defined by differences (including their statistical uncertainties) between data and MC calculations. Using these shape variations, ``toy" MC simulations are used to recalculate the acceptance in each kinematic bin, the maximum variation between the nominal acceptance and the toys being propagated as the systematic uncertainty.

The uncertainty (2.8\% to 5.5\% in \pp and 3.4\% to 6.3\% in \PbPb collisions) in the efficiency of the trigger, muon reconstruction, and muon identification is evaluated bin-by-bin using a data driven technique~\cite{Khachatryan:2010xn}.
Another systematic uncertainty is assigned for the track reconstruction efficiency (4\% per track in \pp collisions~\cite{TRK-11-001} and 6\% in \PbPb collisions~\cite{Khachatryan:2016odn}). This uncertainty, together with all the other listed above as estimated on the \pt integrated bin, are considered as correlated systematic uncertainties. The uncertainties calculated bin-by-bin are considered uncorrelated.
The systematic uncertainty in the cross section measurement is computed as the sum in quadrature of the different contributions mentioned above. The uncertainty of the $\PBpm$~meson decay $\mathcal{B}$ is 3.1\%~\cite{pdg:2016}. The uncertainty of the number of minimum bias events in PbPb, $N_{\text{MB}}$, is 1.0\%. The \TAA uncertainty is $+2.8\%, -3.4\%$~\cite{Khachatryan:2016odn}.

\begin{figure}[tb]
\centering
\includegraphics[width=\cmsFigWidth]{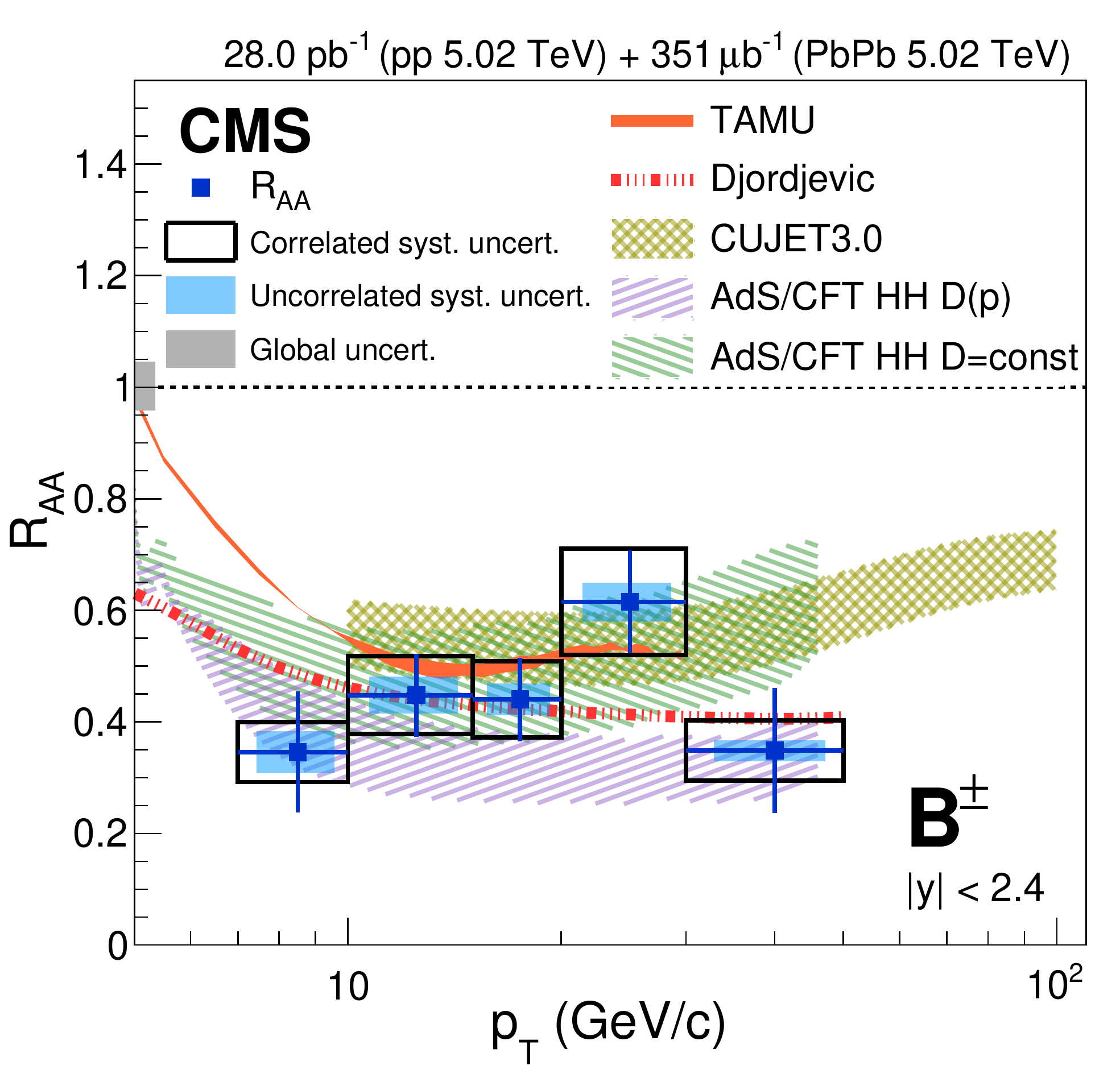}
\caption{The \pt dependence of the nuclear modification factor \RAA of $\PBpm$ measured in \PbPb collisions at $\sqrtsNN=5.02\TeV$. The vertical bars (boxes) correspond to statistical (systematic) uncertainties.
The global systematic uncertainty, represented as a grey box at $\RAA=1$, comprises the uncertainties in the integrated luminosity measurement and \TAA value. Four theoretical calculations are also shown for comparison: TAMU~\cite{He:2011qa, He:2014cla}, Djordjevic~\cite{Djordjevic:2016vfo}, CUJET3.0~\cite{Xu:2015bbz, Xu:2014tda, Xu:2014ica}, and AdS/CFT HH~\cite{Horowitz:2015dta,  AdscftHH}. The line width of the theoretical calculation from Ref.~\cite{He:2011qa, He:2014cla} represents the size of its statistical uncertainty.}
\label{fig:rpaall}
\end{figure}

In Fig.~\ref{fig:crosssections}, the $\pt$-differential production cross sections in \pp and \PbPb collisions measured in the interval $\abs{y} < 2.4$ are presented. The \pp result is compared to the cross section obtained from fixed-order plus next-to-leading logarithm (FONLL) calculations~\cite{FONLLcharmbottomPP3}. The FONLL reference cross section is obtained by scaling the FONLL total b-quark production~\cite{FONLLcharmbottomPP1,FONLLcharmbottomPP2,FONLLcharmbottomPP3} by the world-average production fractions of $\PBp$ of 40.2\%~\cite{pdg:2016}.
The calculated $\PBp$ FONLL reference is consistent with the measured $\PBpm$ \pp spectrum, similarly to what was observed in the previous publications in $\pp$ collisions at $\sqrt{s}= 7$\TeV~\cite{CMSBmesonpp,Chatrchyan:2011pw,Chatrchyan:2011vh,ATLAS:2013cia,LHCb:2013JHEP}.

The nuclear modification factor \RAA, shown in Fig.~\ref{fig:rpaall}, is computed as:
\begin{equation}
 \RAA (\pt) = \frac{1}{\TAA} \frac{{\rd}N^{\PBpm}_\PbPb}{{\rd}\pt} \bigg/ \frac{{\rd}\sigma^{\PBpm}_{\pp}}{{\rd}\pt},
 \label{eq:RAA}
\end{equation}

A clear suppression ($\RAA<1$) of $\PBpm$ mesons production in \PbPb collisions at $\sqrtsNN=5.02\TeV$ is observed. The \RAA is around 0.3 to 0.6 for $\PBpm$ mesons \pt from 7 to 50\GeVc.

The \pt dependence of \RAA is compared to the predictions of: a) two perturbative QCD based models that include both collisional and radiative energy loss (Djordjevic~\cite{Djordjevic:2016vfo}, CUJET3.0~\cite{Xu:2015bbz, Xu:2014tda, Xu:2014ica}); b) a transport theoretical model based on a Langevin equation that includes collisional energy loss and heavy quark diffusion in the medium (TAMU~\cite{He:2011qa, He:2014cla}) and c) a model based on the anti-de-Sitter/conformal field theory correspondence, that includes thermal fluctuations in the energy loss for heavy quarks in a strongly-coupled plasma (AdS/CFT HH~\cite{Horowitz:2015dta, AdscftHH}.) The AdS/CFT HH calculation is provided for two settings of the diffusion coefficient D of the heavy quark propagation through the medium: either dependent on or independent of the quark momentum. The four theoretical calculations differ in several aspects, \eg the modeling of the \PbPb medium (hydrodynamically~\cite{Xu:2015bbz, He:2014cla} or via a Glauber model~\cite{Djordjevic:2016vfo}) and of the energy loss sources (partonic only~\cite{Xu:2015bbz, He:2014cla} or also hadronic~\cite{He:2014cla}), the set of the (nuclear) parton distribution functions used for the initial heavy-quark \pt distributions, \etc Given the current statistical and systematic uncertainties, all these theoretical predictions are roughly compatible with the measurement presented. However, while the present results can not help to resolve the disagreements between different models because of the large uncertainties, including those of the theoretical calculations, they can already be used to optimize parameters settings in such models (\eg the parton-medium coupling parameters in the AdS/CFT model). More precise measurements of the $\PBpm$ mesons \RAA and future results on the angular correlations of $\PBpm$ mesons with other hadrons will allow one to draw a firmer conclusion on the relevance of collisional and radiative processes in the b quark energy loss~\cite{Rohrmoser:2016yct,Cao:2014dja}. The measurement of exclusive $\PBpm$ gives for the first time an unambiguous access to the b hadron quark-flavor content, and represents the first attempt to understand the interactions of beauty and light quarks with each other and with the medium they traverse before hadronization. This lays the groundwork for future measurements of azimuthal asymmetries or relative production ratios like $\PBs/\PBpm$~\cite{He:2012df}.

In summary, the first measurement of the differential production cross section of $\PBpm$ mesons in \pp and \PbPb collisions at $\sqrtsNN=5.02$\TeV has been presented. The $\PBpm$ mesons are measured with the CMS detector at the LHC in the rapidity range $\abs{y}<2.4$ and transverse momentum interval $7<\pt<50\GeVc$ via the reconstruction of one of their exclusive hadronic decay channels, $\Bplusminusdecay$. The nuclear modification factor of $\PBpm$ is measured as a function of its \pt. A strong suppression by about a factor of two is observed in the \PbPb system in comparison to expectations from the scaled \pp reference data. The results are found to be roughly compatible with theoretical calculations incorporating beauty quark diffusion and energy loss in a quark-gluon plasma.

\begin{acknowledgments}
We congratulate our colleagues in the CERN accelerator departments for the excellent performance of the LHC and thank the technical and administrative staffs at CERN and at other CMS institutes for their contributions to the success of the CMS effort. In addition, we gratefully acknowledge the computing centers and personnel of the Worldwide LHC Computing Grid for delivering so effectively the computing infrastructure essential to our analyses. Finally, we acknowledge the enduring support for the construction and operation of the LHC and the CMS detector provided by the following funding agencies: BMWFW and FWF (Austria); FNRS and FWO (Belgium); CNPq, CAPES, FAPERJ, and FAPESP (Brazil); MES (Bulgaria); CERN; CAS, MoST, and NSFC (China); COLCIENCIAS (Colombia); MSES and CSF (Croatia); RPF (Cyprus); SENESCYT (Ecuador); MoER, ERC IUT, and ERDF (Estonia); Academy of Finland, MEC, and HIP (Finland); CEA and CNRS/IN2P3 (France); BMBF, DFG, and HGF (Germany); GSRT (Greece); OTKA and NIH (Hungary); DAE and DST (India); IPM (Iran); SFI (Ireland); INFN (Italy); MSIP and NRF (Republic of Korea); LAS (Lithuania); MOE and UM (Malaysia); BUAP, CINVESTAV, CONACYT, LNS, SEP, and UASLP-FAI (Mexico); MBIE (New Zealand); PAEC (Pakistan); MSHE and NSC (Poland); FCT (Portugal); JINR (Dubna); MON, RosAtom, RAS, RFBR and RAEP (Russia); MESTD (Serbia); SEIDI, CPAN, PCTI and FEDER (Spain); Swiss Funding Agencies (Switzerland); MST (Taipei); ThEPCenter, IPST, STAR, and NSTDA (Thailand); TUBITAK and TAEK (Turkey); NASU and SFFR (Ukraine); STFC (United Kingdom); DOE and NSF (USA).
\end{acknowledgments}

\bibliography{auto_generated}

\cleardoublepage \appendix\section{The CMS Collaboration \label{app:collab}}\begin{sloppypar}\hyphenpenalty=5000\widowpenalty=500\clubpenalty=5000\input{HIN-16-011-authorlist.tex}\end{sloppypar}
\end{document}

%% file: HIN-16-011-authorlist.tex
\textbf{Yerevan Physics Institute,  Yerevan,  Armenia}\\*[0pt]
A.M.~Sirunyan, A.~Tumasyan
\vskip\cmsinstskip
\textbf{Institut f\"{u}r Hochenergiephysik,  Wien,  Austria}\\*[0pt]
W.~Adam, E.~Asilar, T.~Bergauer, J.~Brandstetter, E.~Brondolin, M.~Dragicevic, J.~Er\"{o}, M.~Flechl, M.~Friedl, R.~Fr\"{u}hwirth\cmsAuthorMark{1}, V.M.~Ghete, N.~H\"{o}rmann, J.~Hrubec, M.~Jeitler\cmsAuthorMark{1}, A.~K\"{o}nig, I.~Kr\"{a}tschmer, D.~Liko, T.~Matsushita, I.~Mikulec, D.~Rabady, N.~Rad, H.~Rohringer, J.~Schieck\cmsAuthorMark{1}, J.~Strauss, W.~Waltenberger, C.-E.~Wulz\cmsAuthorMark{1}
% ...
\vskip\cmsinstskip
1:~~Also at Vienna University of Technology, Vienna, Austria\\
% ...
